\documentclass[aps, prd, twocolumn, amsmath, amssymb, superscriptaddress,longbibliography]{revtex4-2}

\usepackage[dvipsnames]{xcolor}
\usepackage{natbib}
\usepackage{graphicx}
\usepackage{enumitem}
\usepackage{siunitx}
\usepackage{booktabs}
\usepackage{multirow}
\usepackage{mathtools}
\usepackage{aas_macros} 

\usepackage[switch]{lineno}

\usepackage[utf8]{inputenc}

\usepackage{comment}

\definecolor{deepgreen}{rgb}{0.2,0.8,0.2}

\definecolor{deepblue}{rgb}{0.2,0.2,0.8}

\definecolor{deepred}{rgb}{0.8,0.2,0.2}

\newcommand{\Eq}[1]{eq.~\ref{#1}}
\newcommand{\Eqs}[1]{eqs.~\ref{#1}}
\newcommand{\Fig}[1]{fig.~\ref{#1}}
\newcommand{\Tab}[1]{table~\ref{#1}}

\newcommand{\vect}[1]{\boldsymbol{\mathbf{#1}}}
\newcommand{\dd}{\mathrm{d}}

\newcommand\myshade{30}
\colorlet{mylinkcolor}{red}
\colorlet{mycitecolor}{orange}
\colorlet{myurlcolor}{orange}

\usepackage[%
  bookmarks=true,
  colorlinks,
  linkcolor=mylinkcolor!\myshade!black,
  urlcolor=myurlcolor!\myshade!black,
  citecolor=mycitecolor!\myshade!black,
  plainpages=false,
  pdfpagelabels,
  final,
  breaklinks=true
]{hyperref}

\hypersetup{
pdftitle={Astrometry with Extended-Path Intensity Correlation},
pdfauthor={Ken Van Tilburg, Masha Baryakhtar, Marios Galanis, Neal Weiner},
pdfkeywords={intensity interferometry, astrometry}
}

\begin{document}
\title[Astrometry with Extended-Path Intensity Correlation]{Astrometry with Extended-Path Intensity Correlation}

\author{Ken Van Tilburg}
\email{kenvt@nyu.edu | kvantilburg@flatironinstitute.org}
\affiliation{Center for Cosmology and Particle Physics, Department of Physics, New York University,
New York, NY 10003, USA}
\affiliation{Center for Computational Astrophysics, Flatiron Institute, New York, NY 10010, USA}

\author{Masha Baryakhtar}
\email{mbaryakh@uw.edu}
\affiliation{Department of Physics, University of Washington, Seattle WA 98195, USA}

\author{Marios Galanis}
\email{mgalanis@perimeterinstitute.ca}
\affiliation{Perimeter Institute for Theoretical Physics, Waterloo, Ontario N2L 2Y5, Canada}

\author{Neal Weiner}
\email{neal.weiner@nyu.edu}
\affiliation{Center for Cosmology and Particle Physics, Department of Physics, New York University,
New York, NY 10003, USA}

\date{\today}

\begin{abstract}
  Intensity interferometry---the correlation of spatially separated light intensities---has historically been an important tool for precision optical astronomical observations. 
  However, due to the extremely narrow field of view, its scope has been limited to studies of the morphology of very bright emission regions, primarily determinations of  angular diameters of nearby hot stars.
  We propose adding an adjustable path extension into the detector optics which creates a primary interference fringe for widely separated sources, allowing maximum source separations parametrically larger than the angular resolution. 
  This \emph{extended-path intensity correlator} (EPIC), augmented with advances in single-photon detectors and spectroscopic gratings, would enable ground-based astrometry at microarcsecond-level precision in a field of view as large as several arcseconds. 
  EPIC has the potential to revolutionize astrophysical and cosmological observations requiring high-precision differential astrometry on sources of high surface brightness.
  We outline how EPIC can be employed to detect the astrometric wobble of Earth-like planets around Sun-like stars at tens to hundreds of parsecs, and
  expect that EPIC's larger field of view will expand the power of intensity interferometry to a broad range of astronomical applications.
\end{abstract}

\maketitle

\subsection*{Introduction}
Interferometry---the precision measurement of phase differences between paths---has a long history of revolutionary advances in physics and astronomy~\cite{Michelson1887On}. In the last decade alone, for instance, amplitude interferometry has led to ground-breaking observations of gravitational waves~\cite{PhysRevLett.116.061102} and images of light rings~\cite{Akiyama_2019} and orbits~\cite{2018A&A...618L..10G} near black hole horizons.
{\it  Intensity interferometry}, pioneered by Hanbury Brown and Twiss~\cite{1954PMag...45..663B,1956Natur.177...27B,1956Natur.178.1046H}, utilizes second-order coherence of light, by correlating \emph{intensities} instead of amplitudes at two separated telescopes. The technique enables exceptional angular resolution scaling as the inverse of the telescope separation, which can be made arbitrarily large since the (optical) light need not be physically recombined as in an amplitude interferometer. The method primarily requires fast photon counting to precisely measure intensity as a function of time and large light collection areas to tease out the small statistical correlations in photon arrival times, and is robust under poor atmospheric conditions~\cite{1974iiaa.book.....B}. 
One of the fundamental limitations of intensity interferometry is that correlations diminish dramatically on angular scales large compared to the resolution, restricting this technique to measurements of  stellar angular diameters~\cite{1967MNRAS.137..393H,1974MNRAS.167..121H,2020NatAs...4.1164A,2021MNRAS.506.1585Z} and close binary orbits~\cite{1971MNRAS.151..161H,2018MNRAS.480..245G}. A novel approach is needed to broaden the scope of intensity interferometry, literally and figuratively. We propose such an idea here.

\begin{figure}[t]
\centering
\includegraphics[width=1\columnwidth, trim = 0 0 0 0]{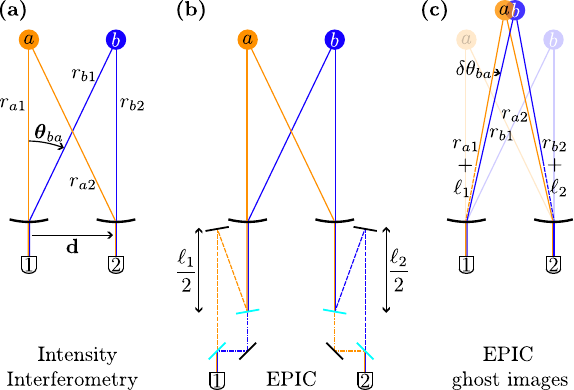}
\caption{
\textbf{(a)} Intensity interferometry with two telescopes $p=1,2$, separated by baseline $\vect{d}$, of two sources $s=a,b$ with relative angle $\vect{\theta}_{ba}$ and distances $r_{sp}$. \textbf{(b)} Extended-Path Intensity Correlation (EPIC): (a) with additional $50/50$ beamsplitters, creating internal interferometers with unequal path lengths. The two-photon amplitude depicted (in orange and dark blue) extends the path $a\to 1$ ($b \to 2)$ by $\ell_1$ ($\ell_2$) relative to $b \to 1$ ($a \to 2$), permitting operation near the main fringe even for large $\theta_{ba}$. \textbf{(c)} The EPIC fringe and distances from (b) are equivalent to a ghost image pair of $a$ and $b$ separated by an arbitrarily small angle $\delta \theta_{ba}$.}\label{fig:basics}
\end{figure}

\subsection*{Astrometry with Intensity Interferometry}
In two-source intensity interferometry, the primary observable is the correlation between light \emph{intensities} from two sources $a$ and $b$ separated by angle $\vect{\theta}_{ba} \equiv \hat{\vect{\theta}}_b - \hat{\vect{\theta}}_a$ at two detectors $1$ and $2$ with baseline $\vect{d}$ (\Fig{fig:basics}a). The intensity fluctuations are positively correlated when $\vect{\theta}_{ba} \cdot \vect{d} \lesssim \lambda/2$, where $\lambda$ is the wavelength of the recorded light. For larger baselines, the intensity correlation exhibits fringes for $d$ at integer multiples of $\vect{\theta}_{ba} \cdot \hat{\vect{d}}/\lambda$. For two equally bright, nearly monochromatic point sources with mean intensity $I_0$, the intensities $I_{1,2}$ at the two telescopes are:
\begin{alignat}{3}
  \! \! \frac{I_1(t_1)}{I_0} &=  1 + \cos \big[ k (r_{a1} - r_{b1}) + \phi_{a} \! \left(t^\mathrm{ret}_{a1} \right) \! - \phi_{b} \! \left(t^\mathrm{ret}_{b1}\right) \! \big], \nonumber \\
  \! \! \frac{I_2(t_2)}{I_0} &=  1 + \cos \big[ k (r_{a2} - r_{b2}) + \phi_{a} \! \left(t^\mathrm{ret}_{a2} \right) \! - \phi_{b} \! \left(t^\mathrm{ret}_{b2}\right) \! \big]; \label{eq:I12} 
\end{alignat}
where $k = 2\pi / \lambda$ is the wavenumber of the light.

In this idealized classical model, the phases $\phi_s$ fluctuate randomly as a function of the retarded time $t_{sp}^\mathrm{ret} = t_p - r_{sp}/c$ from the telescope $p=1,2$ to the source $s = a,b$ at a distance $r_{sp}$. For a relative time delay $\tau \equiv t_2-t_1$ equal to $(r_{a2} - r_{a1})/c$, the phase $\phi_a$ and the intensity fluctuations from source $a$ will be (positively) correlated at both telescopes. If source $b$ is sufficiently close in angle to $a$, then the same choice of $\tau$ will simultaneously lead to nearly equal $\phi_b$, generating positive correlations for both phases and thus excess fractional intensity correlation:
\begin{alignat}{2}
  C(\vect{d},\tau) &\equiv \frac{\langle I_1(t) I_2(t+\tau) \rangle}{\langle I_1 \rangle \langle I_2 \rangle} - 1 = \frac{1 + \cos \left( k \vect{d} \cdot \vect{\theta}_{ba} \right)}{2}. \label{eq:C_1}
\end{alignat}
Brackets $\langle  \cdot \rangle$ signify phase averaging over $\phi_{a,b}$, and we use the small-angle approximation: 
\begin{alignat}{2}
  \big(r_{b1} + r_{a2}\big) - \big(r_{a1} + r_{b2}\big)  = \vect{\theta}_{ba} \cdot \vect{d}. \label{eq:path_diff}
\end{alignat}
That is, the ``crossed'' paths in \Fig{fig:basics}a are longer than the ``uncrossed'' paths by the relative angular source separation times the baseline distance. Classically, this information is carried in the relative phases in the four-point correlation of the electromagnetic field (\Eq{eq:C_1}). Quantum mechanically, one can view this as the two-photon amplitudes $\langle 1{,}2 \vert b{,}a \rangle$ and $\langle 1{,}2 \vert a{,}b \rangle$ interfering with a relative propagation phase $k \vect{d} \cdot \vect{\theta}_{ba}$~\cite{mandel1995optical}.

Measurement of the intensity correlations of \Eq{eq:C_1} yields the relative source separation $\vect{\theta}_{ba}$ with a fiducial angular resolution:
\begin{alignat}{2}
  \sigma_{\theta_\mathrm{res}} = \frac{1}{k d} = \frac{\lambda}{2 \pi d} \approx 1.64 \, \mathrm{\mu as}  \left(\frac{\lambda}{500\,\mathrm{nm}} \right) \! \! \left( \frac{10 \, \mathrm{km}}{d}\right) \label{eq:theta_res}
\end{alignat}
along the direction $\hat{\vect{d}}$. In practice, by {recording the arrival times of photons}, one can construct an estimator for the instantaneous intensities $I_{1,2}$ and their excess fractional correlation $C(\vect{d},\tau)$~\cite{longpaper}. A positive value of the latter is a direct measure of ``photon bunching'', the intuitively surprising result that near-simultaneous photon arrival times (after applying an appropriate time delay $\tau$) are more likely to occur than from random chance~\cite{mandel1995optical}. An inversion of the function in \Eq{eq:C_1} yields a multivalued map $C(\vect{d},\tau) \mapsto k \vect{\theta}_{ba} \cdot \vect{d}$, from which the relative separation $\vect{\theta}_{ba}$ between the light centroids of $a$ and $b$ can be measured with a precision of $\sigma_{\delta \theta} \sim \sigma_{\theta_\mathrm{res}} / \mathrm{SNR}$ (\Eq{eq:sigma_theta}), where $\mathrm{SNR}$ is the total signal-to-noise ratio on the $C(\vect{d},\tau)$ observation (Methods~\ref{app:obs},\ref{app:centroid}). The degeneracy of the multivalued  map from correlator to separation can be broken---$\vect{\theta}_{ba}$ can be assigned to a unique fringe---by observing the intensity correlations in many spectral channels (each with different $k$) and as a function of time, since the projected baseline $\hat{\vect{\theta}}_{ba} \cdot \vect{d}$ changes (primarily) due to Earth's rotation.

Ground-based differential astrometry with a fiducial resolution of \Eq{eq:theta_res} and even more astonishing light-centroiding precision opens up a myriad of scientific applications, but traditional intensity interferometry is severely hamstrung by two problems: an extremely limited field of view (FOV) and low SNR. Our proposal of Extended-Path Intensity Correlation (EPIC) solves the former, while multichannel observations and recent technological improvements in ultrafast single photon detection can ameliorate the latter~\cite{longpaper,2012NewAR..56..143D,2014JKAS...47..235T,2022OJAp....5E..16S,Chen:2022ccn,2022AJ....163...92H}. 

The small-FOV limitation arises from the finite bandwidth of the detected light. In each spectral channel of spectral resolution $\mathcal{R} \equiv k / \sigma_k$ with Gaussian spread $\sigma_k$ around wavenumber $k$, bandwidth smearing leads to a loss of fringe contrast for $\left | \sigma_k \vect{\theta}_{ba} \cdot \vect{d} \right | \gtrsim 1$ or an angular dynamic range
\begin{align}
  \! \! \sigma_{\Delta \theta} 
  = \frac{\sqrt{2}}{\sigma_k d} 
  \approx 12 \,\text{mas}  \left(\frac{\mathcal{R}}{5{,}000}\right) \! \! \left(\frac{\lambda}{500\,\text{nm}}\right) \! \! \left(\frac{10\,\text{km}}{d}\right) \! , \label{eq:Delta_theta}
\end{align}
analogous to the ``coherent FOV'' of amplitude interferometers. In other words, the source separation for which an intensity interferometer produces sharp fringes, as in \Eq{eq:C_1}, has to be less than $\mathcal{R}$ times the resolution,~$\theta_{ba} \lesssim \mathcal{R} \sigma_{\theta_\mathrm{res}}$. This is a serious impediment if one desires microarcsecond-level angular resolution for sources separated by arcseconds. Such a high spectral resolution with dense coverage over a wide spectral range is unachievable by ground-based telescopes, and operation at a very high-order fringe would impose prohibitive requirements on fringe stability and possibly lead to fringe confusion. Furthermore, at separations for which $\vect{\theta}_{ba} \cdot \vect{d} / c$ is larger than the relative timing resolution $\sigma_t$ (typically longer than the coherence time $1/c\sigma_k$ of the light in each spectral channel), a total loss of mutual second-order coherence occurs, since the wavefronts from sources $a$ and $b$ arrive at the telescopes at different relative times $\tau$. The timing precision defines the angular scale
\begin{alignat}{2}
    \sigma_{\hat{\vect{\theta}}} = \frac{2 \sigma_t}{d} \approx 124\,\text{mas}  \left(\frac{\sigma_t}{10\,\mathrm{ps}}\right) \! \! \left( \frac{10\,\text{km}}{d}\right) \! , \label{eq:hat_theta}
\end{alignat}
at which \emph{global} astrometry is possible with intensity interferometry, assuming $\sigma_t > 1/c\sigma_k$.

\subsection*{Extended-Path Intensity Correlation}
In this work, we propose a variant of intensity interferometry that parametrically decouples the maximum source separation from the angular resolution, effectively increasing the field of view by orders of magnitude while retaining its light-centroiding precision. 
To ``point'' an interferometer at a target of interest, a relative time delay $\tau$ can be applied offline, but any \emph{detector}-dependent phase shift 
cannot point at two targets at once, since it will contribute to both terms in brackets on the LHS of \Eq{eq:path_diff}, thus leaving the RHS unchanged. A \emph{source}-dependent phase shift has to be added in real time, in the telescope optics, to lengthen e.g.~$r_{a1}$ and/or $r_{b2}$ without affecting $r_{b1}$ nor $r_{a2}$. We refer to intensity interferometry with this additional shift as ``Extended-Path Intensity Correlation'' (EPIC). Outside the context of optical astronomy, similar approaches have been proposed for gravitational-wave detection~\cite{1999PhRvD..59j2003T} and tests of quantum mechanics~\cite{1989PhRvL..62.2205F}.

\begin{figure}[t!]
  \centering
  \includegraphics[width=0.5\columnwidth, trim = 0 0 0 00, clip]{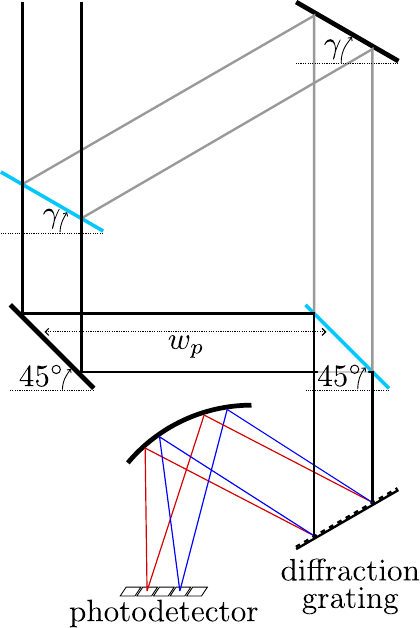}
  \caption{The path extension stage of an EPIC telescope accepts collimated light into its beamsplitter at an angle $\gamma$. The reflected path (gray) is longer than that of the transmitted path (black) by $\ell_p = w_p (\csc 2\gamma +\cot 2 \gamma - 1)$. At one (or both) of the output ports of the beam recombiner, the light is spectrally split by a reflective diffraction grating and focused onto a single-photon detector array whose pixels constitute the spectral channels.}\label{fig:optics}
\end{figure}

In EPIC, the light from both sources enters the same telescope aperture and is equally split into two paths of different lengths (with difference $\ell_p$) before it is recombined into one beam in a Mach-Zehnder geometry (\Fig{fig:optics}). The probability for the light from source $s$ to be detected by each telescope's photodetector $p$ is the superposition of two possible amplitudes with respective path lengths:
\begin{alignat}{2}
  r_{sp} \quad \text{and} \quad r_{sp} + \ell_p \qquad (s = a,b; ~ p = 1,2). \label{eq:path_extension}
\end{alignat}
There are $2^4 = 16$ propagation path combinations contributing to the intensity correlator $\langle I_1 I_2 \rangle$, corresponding to the 4 independent possibilities in \Eq{eq:path_extension}.

One possible fringe choice is the one where only light from $a \to 1$ is extended by $\ell_1$, and that of $b \to 2$ by $\ell_2$ (\Fig{fig:basics}b), leading to a modification of \Eq{eq:path_diff} of the doubly-differential propagation path:
\begin{alignat}{2}
  &\big[r_{b1} + r_{a2}\big] - \big[(r_{a1} + \ell_1) + (r_{b2} + \ell_2) \big] \nonumber \\
  & \hspace{9.5em} = \vect{\theta}_{ba} \cdot \vect{d} - (\ell_1 + \ell_2) \equiv \delta \theta_{ba}. \label{eq:path_epic}
\end{alignat}
The path difference and $\delta \theta_{ba} \equiv (\vect{\theta}_{ba} - \vect{\theta}_{ba}^\mathrm{ref}) \cdot \hat{\vect{d}}$ can thus be made arbitrarily small by adjusting the reference angle $\vect{\theta}_{ba}^\mathrm{ref} \equiv \hat{\vect{d}} (\ell_1 + \ell_2) / d$ close to the true separation $\vect{\theta}_{ba}$. 

The fringe of \Eq{eq:path_epic} can be selected (i.e.~the other fringes ignored) by picking the time delay $\tau$ equal to the optimal value $\tau^\mathrm{opt} = - (\hat{\vect{\theta}}_a + \hat{\vect{\theta}}_b) \cdot \vect{d}/2 + \ell_2 - \ell_1$. The excess fractional intensity correlation in EPIC is:
\begin{alignat}{2}
  C(\vect{d},\tau^\mathrm{opt}) &\simeq
  \frac{1}{4\sqrt{2} c\sigma_k \sigma_t}  \Bigg \lbrace 
    \frac{\langle I_a \rangle^2 +  \langle I_b \rangle^2}{\left(\langle I_a \rangle +  \langle I_b \rangle\right)^2} 
      \exp\bigg[\frac{-(\delta \theta_{ba})^2}{2 \sigma_{\hat{\vect{\theta}}}^2}\bigg]  \label{eq:C_epic} \\
    &+ \frac{2 \langle I_a \rangle \langle I_b \rangle}{\left(\langle I_a \rangle +  \langle I_b \rangle\right)^2}
      \cos\left[\frac{\delta\theta_{ba}}{\sigma_{\theta_\mathrm{res}}}  \right] 
      \exp\bigg[\frac{-(\delta \theta_{ba})^2}{2 \sigma_{\Delta \theta}^2} \bigg] \nonumber
  \Bigg \rbrace,
\end{alignat}
with $\sigma_{\theta_\mathrm{res}}$, $\sigma_{\Delta \theta}$, and $\sigma_{\hat{\vect{\theta}}}$ from \Eqs{eq:theta_res},~\ref{eq:Delta_theta}, and \ref{eq:hat_theta}. We include effects from unequal source fluxes $\langle I_a \rangle \neq \langle I_b \rangle$, the overall fringe contrast suppression due to a timing resolution $\sigma_t$, and smearing over the bandwidth $\sigma_k$~\cite{longpaper}. 
Equation~\ref{eq:C_epic} shows that the angular dynamic range $\sigma_{\Delta \theta}$ is \emph{not} enhanced, but that the path extensions create ``ghost images'' of the sources, as if they are only displaced by a small angle $\vect{\theta}_{ba} - \vect{\theta}_{ba}^\mathrm{ref}$ (\Fig{fig:basics}c) near the main fringe. Since $\vect{\theta}_{ba}^\mathrm{ref}$ is known, the source separation can be measured via inversion of the map in \Eq{eq:C_epic}. 

\begin{table}[t!]
  \centering
  \renewcommand{\arraystretch}{1.2} 
  \begin{tabular}{c | r r r r | r r r}
  \hline \hline
Phase & \multicolumn{1}{c}{$D$} & \multicolumn{1}{c}{$\sigma_t$} & \multicolumn{1}{c}{$\mathcal{R}$} & \multicolumn{1}{c|}{$n_\mathrm{arr}$} & \multicolumn{1}{c}{$\sigma_{\Delta \theta}$} & \multicolumn{1}{c}{$\sigma_{\hat{\vect{\theta}}}$}  & \multicolumn{1}{c}{$\sigma_{\delta \theta}$} \\
\hline
  I   & {$4\,\mathrm{m}$}   & {$30\,\mathrm{ps}$}   & $5{,}000$   & 1   & $\SI{0.16}{\arcsecond}$ & $\SI{5.2}{\arcsecond}$    & $22 \, \mathrm{\mu as}$  \\
  II  & {$10\,\mathrm{m}$}  & {$10\,\mathrm{ps}$}   & $10{,}000$  & 1   & $\SI{0.33}{\arcsecond}$ & $\SI{1.8}{\arcsecond}$    & $1.5  \, \mathrm{\mu as}$ \\
  III & {$10\,\mathrm{m}$}  & {$3\,\mathrm{ps}$}    & $20{,}000$  & 10  & $\SI{0.66}{\arcsecond}$ & $\SI{0.52}{\arcsecond}$   & $0.056 \, \mathrm{\mu as}$  \\
  \hline\hline
  \end{tabular}
  \caption{EPIC  parameters for Phases I/II/III: aperture diameter $D$, timing resolution $\sigma_t$, spectral resolution $\mathcal{R} = k/\sigma_k$, and number of detectors per array site $n_\mathrm{arr}$. Also shown are the resulting angular dynamic range $\sigma_{\Delta \theta}$ (\Eq{eq:Delta_theta}), global astrometric resolution $\sigma_{\hat{\vect{\theta}}}$ (\Eq{eq:hat_theta}), and light-centroiding precision $\sigma_{\delta \theta}$ after an observation time $t_\mathrm{obs} = 10^4 \,\mathrm{s}$ for a pair of Sun-like stars at a distance of $100\,\mathrm{pc}$. For such a source pair, the optimal (projected) baseline distance is $d = 0.71\,\mathrm{km}$ for a fiducial angular resolution of $\sigma_{\theta_\mathrm{res}} = 23 \, \mathrm{\mu as}$ at $\lambda = 500 \, \mathrm{nm}$ (\Eq{eq:theta_res}). We assume a photodetection efficiency of $\eta = 0.5$ in addition to irreducible intensity decrease in the path extension stage, and that unpolarized light between $\lambda = 300\, \mathrm{nm}$ and $\lambda = 1{,}000 \, \mathrm{nm}$ is recorded in spectral channels separated by factors of $e^{2/\mathcal{R}}$. See Methods~\ref{app:centroid} and Ref.~\cite{longpaper} for details.}
\label{tab:phases}
\end{table}

\begin{figure*}[t!]
  \centering
  \includegraphics[width=0.7\textwidth, trim = 0 0 0 0]{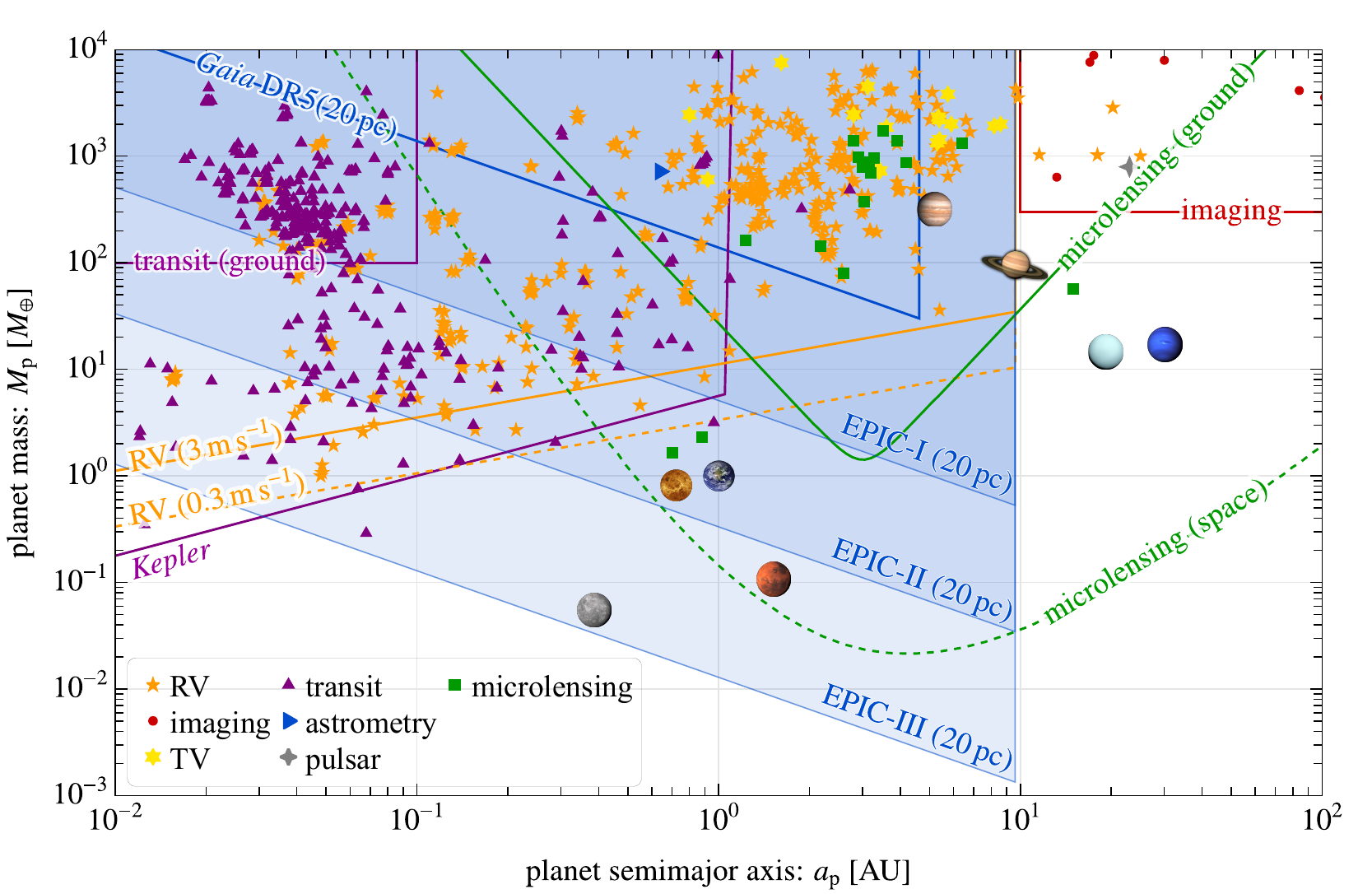}
  \caption{Projected EPIC sensitivity to exoplanets around Sun-like stars (adapted from Ref.~\cite{2021ARA&A..59..291Z}). Detection sensitivity at $3\sigma_{\Delta \theta_\mathrm{h}}$ is shown as a function of semimajor axis $a_\text{p}$ and mass $M_\text{p}$ by the blue regions for EPIC Phase $\lbrace \text{I,II,III} \rbrace$ with wobble precision of $\sigma_{\Delta \theta_\mathrm{h}} = \lbrace 260,17,0.65 \rbrace \times 10^{-3} \mathrm{\mu as}$. The reach of other detection techniques with current (future) capabilities is shown in solid (dashed) lines: the astrometric sensitivity of \textit{Gaia} DR5 (blue)~\cite{2016A&A...595A...1G}, radial velocity (RV) surveys (orange)~\cite{2016PASP..128f6001F}, transit methods (purple)~\cite{2010Sci...327..977B,2006PASP..118.1407P,2004PASP..116..266B}, direct imaging (red)~\cite{2016PASP..128j2001B}, and microlensing (green)~\cite{2019ApJS..241....3P}. Confirmed exoplanets in multiple-star systems from the NASA Exoplanet Archive~\cite{*[{}] [{, \url{https://exoplanetarchive.ipac.caltech.edu/}, [acquired May 2023].}] 2013PASP..125..989A} discovered through these methods and eclipse timing variations (TV) are overlaid, as well as the Solar System planets (image credit: NASA).} \label{fig:exoplanets}
\end{figure*}

\subsection*{EPIC Sensitivity and Maximum Separation}
We anticipate an EPIC program to develop in three Phases that would rapidly reach unprecedented light-centroiding precision on bright stars;  the benchmark parameters and expected performance are listed in \Tab{tab:phases}.

Fractional intensity correlations manifest as coincident photon detections. An optimal estimator for $C$ (\Eq{eq:C_epic}) has variance $\sigma_C^2 = t_\mathrm{obs} / (\sqrt{4\pi} \sigma_t N_1 N_2)$ where $N_p = t_\mathrm{obs} \langle I_p \rangle \eta_p A_p / (\hbar c k)$ is the expected number of photons at telescope $p$ per spectral channel centered on wavenumber $k$ after an observation time $t_\mathrm{obs}$, with efficiency $\eta_p$ and aperture area $A_p$ (which can be enhanced by $n_\mathrm{arr}$ telescopes per array site)~\cite{longpaper}.
The total SNR is the quadrature sum of $C/\sigma_C$ over all channels, which can be logarithmically spaced by factors of $e^{2/\mathcal{R}}$~(Methods~\ref{app:centroid}). The SNR is halved for unpolarized light. Spectral resolutions of $\mathcal{R} \geq 5{,}000$ are standard with commercially available diffraction gratings~\cite{eversberg2015spectroscopic}, while timing resolutions approaching $\sigma_t \lesssim 3\,\mathrm{ps}~(30\,\mathrm{ps})$ have been achieved with superconducting nanowire single photon detectors~\cite{3pssnspd} (single photon avalanche diodes~\cite{Zappa2018,becker2022}).

EPIC can perform high-precision measurements at source separations orders of magnitude larger than traditional intensity interferometry. Instead of being limited by spectral or timing resolution, the maximum source separation is now set by refractive phase errors from the turbulent atmosphere.  These phase fluctuations become important at opening angles greater than the isoplanatic angle $\theta_0$, of the order of a few arcseconds~\cite{2002ESOC...58..321S},  yielding a suppression in the correlation by a factor $\exp \lbrace-(\theta_{ba}/\theta_0)^{5/3}\rbrace$ in the second line of \Eq{eq:C_epic}  (Methods~\ref{app:atm})~\cite{longpaper}.  
For separations of order the isoplanatic angle, the main EPIC fringe ($\delta \theta_{ba} \simeq 0$) is obtained with path extensions of $\ell_1 + \ell_2 \approx 4.8 \, \mathrm{cm} \, (\vect{\theta}_{ba} \cdot \vect{d}) / (1 \, \mathrm{arcsec} \cdot 10 \, \mathrm{km})$. 

\subsection*{Applications: Exoplanet Detection}
High-precision differential astrometry benefits many scientific applications~\cite{2021ARA&A..59...59B}, including binary-orbit characterization~\cite{1971MNRAS.151..161H,longpaper}, gravitational microlensing of stars~\cite{1986ApJ...304....1P,longpaper} and quasars~\cite{2004A&A...416...19T,companionquasar2023}, galactic dynamics~\cite{2020ARA&A..58..205H,2021PhRvL.127x1104B,longpaper}, and orbits around Sagittarius A*~\cite{2009ApJ...692.1075G,2008ApJ...689.1044G,companionsaga2023}. Here, we focus on exoplanet detection to illustrate EPIC's capabilities.

The gravitational pull of an orbiting exoplanet causes a small periodic wobble in its host star's position. Dozens of exoplanets have been discovered astrometrically with amplitude interferometers~\cite{2010AJ....140.1657M,2014A&A...565A..20S} and \textit{Gaia}~\cite{2023A&A...674A..34G}; thousands more are expected soon~\cite{2014ApJ...797...14P}. The challenge is the small amplitude of the astrometric wobble: $\Delta \theta_\mathrm{h} = (M_\mathrm{p}/M_\mathrm{h})(a_\mathrm{p}/D_\mathrm{h}) \approx 0.15 \, \mathrm{\mu as}$ for exoplanet mass $M_\mathrm{p} = M_\oplus$, host mass $M_\mathrm{h} = M_\odot$, circular orbit's semimajor axis $a_\mathrm{p} = \mathrm{AU}$, and line-of-sight distance $D_\mathrm{h} = 20 \, \mathrm{pc}$. \textit{Gaia}'s final-mission wobble light-centroiding precision will be $\sigma_{\Delta \theta_\mathrm{h}} \approx 7\,\mathrm{\mu as}$ for typical nearby stars, limiting its sensitivity to massive exoplanets.

EPIC can greatly increase the discovery potential for Earth-mass exoplanets around host stars with a nearby reference source---either in multiple-star systems or accidental doubles. The per-epoch light-centroiding precision of EPIC Phases $\lbrace \mathrm{I}, \mathrm{II}, \mathrm{III} \rbrace$ is $\sigma_{\delta \theta} \approx \lbrace 4.5, 0.29, 0.011\rbrace  \, \mathrm{\mu as}$ for a pair of Sun-like stars at $D_\mathrm{h} = 20 \, \mathrm{pc}$ (scaling as $\sigma_{\delta \theta} \propto D_\mathrm{h}$). After $N_\mathrm{obs} = 300$ observations over $30\,\mathrm{yr}$, wobble precisions of $\sigma_{\Delta \theta_\mathrm{h}} = \sigma_{\delta \theta} / \sqrt{N_\mathrm{obs}}$ enable detection of Earth-Sun-like systems with a nearby reference star at distances up to $20\,\mathrm{pc}$ ($400\,\mathrm{pc}$) at $3\sigma$ with EPIC-II(III). 
The exoplanet parameter space accessible to EPIC astrometry (blue regions in \Fig{fig:exoplanets}) is complementary to that of other techniques. Transits (purple) and radial-velocity signatures (orange) are most sensitive to exoplanets at small semimajor axes, while direct imaging (red) favors large planets far away from their host star. Microlensing (green) due to chance alignments of exoplanetary systems with background stars can lead to detection of very low-mass systems but rapidly loses sensitivity for small orbits. In regions where EPIC shares sensitivity with other techniques, the respective observational biases would be different, aiding population synthesis analyses over a wider range of systems~\cite{2021ARA&A..59..291Z}.

\subsection*{Conclusion}
Intensity interferometry holds the promise of exceptional angular resolution on bright sources, but has been hampered by its narrow FOV in its uses for differential astrometry. By introducing variable, source-dependent path extensions, EPIC enlarges the observable source separation to the maximum allowed by atmospheric disturbances. Combined with advances in spectroscopy and fast single-photon detection, EPIC's differential light-centroiding performance will facilitate new exoplanet discoveries and unlock many other scientific applications benefiting from narrow-angle astrometry.

  
\subsubsection*{Acknowledgments}
We thank Gordon Baym, Megan Bedell, Karl Berggren, Michael Blanton, Matteo Cantiello, Calvin Chen, Cyril Creque-Sarbinowski, Liang Dai, Neal Dalal, Julianne Dalcanton, David Dunsky, Peter Graham, David Hogg, Marius Kongsore, Miguel Morales, Oren Slone, and David Spergel for valuable conversations and input.
KVT thanks Jason Aufdenberg, Matthew Brown, James Buckley, Dainis Dravins, David Kieda, Michael Lisa, Nolan Matthews, Andrei Nomerotski, Ue-Li Pen, Naomi Vogel, Shiang-Yu Wang, and Luca Zampieri for fruitful conversations, and Sebastian Karl for pointing out Ref.~\cite{1989PhRvL..62.2205F},  during the 2023 Workshop on Stellar Intensity Interferometry at The Ohio State University.

KVT is supported by the National Science Foundation under Grant PHY-2210551. 
MB is supported by the DOE Office of Science under Award Number DE- SC0022348, and the Royal Research Fund, the Department of Physics, and the College of Arts and Science at the University of Washington. 
NW is supported by NSF under award PHY-1915409, by the BSF under grant 2018140, and by the Simons Foundation. 

The Center for Computational Astrophysics at the Flatiron Institute is supported by the Simons Foundation. 
Research at Perimeter Institute is supported in part by the Government of Canada through the Department of Innovation, Science and Economic Development and by the Province of Ontario through the Ministry of Colleges and Universities.
MB, MG, and KVT thank the Institute for Nuclear Theory at the University of Washington for its kind hospitality and stimulating research environment. The  INT is supported in part by the U.S. Department of Energy grant No.~DE-FG02-00ER41132. 
This work was performed in part at the Aspen Center for Physics, which is supported by National Science Foundation grant PHY-1607611 and PHY-2210452. The participation of MB at the Aspen Center for Physics was supported by the Simons Foundation.

This research has made use of the NASA Exoplanet Archive, which is operated by the California Institute of Technology, under contract with the National Aeronautics and Space Administration under the Exoplanet Exploration Program.

\vspace{3em}

\bibliography{EPIC_short} 


\vspace{3em}

{\noindent\bfseries\Large{Methods}}

\section{Observational Procedure} \label{app:obs}
EPIC is ideally suited for differential astrometry on relatively bright sources with apparent magnitude $m \lesssim 15$, especially with Phases II and III. The reference sources in \Tab{tab:phases}, a pair of Sun-like stars at $100\,\mathrm{pc}$, have $m \sim 10$. They are roughly at the limiting magnitude of Phase I, which just about reaches a single-epoch SNR of order unity, as $\sigma_{\delta \theta} \approx \sigma_{\theta_\mathrm{res}}$ after $t_\mathrm{obs} = 10^4 \, \mathrm{s}$. Statistically, most candidate source pairs on which EPIC can be applied will be separated by an angle of order the maximal one, the isoplanatic angle. This is significantly larger than the seeing angle of ground-based observatories and the diffraction limit of space-based telescopes such as \textit{HST}, \textit{JWST}, and \textit{Gaia}, thus allowing identification and characterization \emph{prior} to EPIC observations.

For the bright source pairs under consideration, \textit{Gaia} will be able to provide astrometry with $\mathcal{O}(20 \, \mathrm{\mu as})$ accuracy across the full sky. For the first EPIC observations, one would choose path extensions $\ell_1(t)+\ell_2(t)=\vect{\theta}_{ba}^\mathrm{ref}(t)\cdot \vect{d}(t)$ for a reference angle close to the \textit{Gaia} value and commensurate with the time dependence of Earth's rotation and the relative proper motion and parallax of the sources. For Phase I, \textit{Gaia}'s accuracy should be sufficient to place the source on the primary EPIC fringe, as the optimal baseline corresponds to an angular resolution worse than $20\,\mathrm{\mu as}$ (cfr.~\Tab{tab:phases} and Methods~\ref{app:centroid}). In Phases II \& III, there may initially be fringe confusion, but with sufficient SNR across different spectral channels and at different times, this degeneracy can be broken. Subsequent EPIC observations can then use the updated light centroid in the adjustment of $\vect{\theta}_{ba}^\mathrm{ref}$~\cite{longpaper}.

For practical simplicity and computational efficiency, we envision observations broken into a series of short intervals of $t_\mathrm{obs} \sim 10^2 \, \mathrm{s}$, discretely varying $\ell_1(t)+\ell_2(t)$ and computing the intensity correlation for each spectral channel and time interval individually. For each generalized bin of wavenumber $k$ and time interval $t$, the correlation $C$ may be below statistical noise levels. However, a global fit to these binned data can extract $\delta \theta_{ba} = (\vect{\theta_{ba}}-\vect{\theta}_{ba}^\mathrm{ref}) \cdot \hat{\vect{d}}$ through the wavenumber and time dependence as the argument of the correlator in \Eq{eq:C_epic} gradually transits across the interference pattern, with a precision given by \Eqs{eq:sigma_channel} and~\ref{eq:sigma_theta} (Methods~\ref{app:centroid}). To achieve the target light-centroiding precision, the path extensions $\ell_{1,2}$ need to be measured and controlled at the sub-wavelength level; we perform a detailed study of tolerances and aberrations in follow-up work~\cite{longpaper}.

\section{Light-Centroiding Precision} \label{app:centroid}
The differential light-centroiding precision $\sigma_{\delta \theta}$ depends on the spectrum, surface brightness, and angular size of the sources $a$ and $b$. The fiducial angular resolution of \Eq{eq:theta_res} can be made arbitrarily small by taking $d \to \infty$, but light-centroiding precision suffers in this limit because fringe contrast is lost due to form factor suppression of the sources' finite angular sizes. Here, we outline the calculation of the \emph{optimal} baseline and resolution, and of the resulting light-centroiding precision used in \Tab{tab:phases} and \Fig{fig:exoplanets}.

We model stars, the source targets of primary interest, as circular disks of uniform temperature $T_s$ and angular radius $\theta_s = R_s / D_s$ where $R_s$ is the physical radius and $D_s$ the line-of-sight distance of the star. The mean light intensity in a spectral channel centered at $k$ with Gaussian standard deviation $\sigma_k = k / \mathcal{R}$ is then:
\begin{alignat}{2}
  \langle I_s \rangle = \frac{\hbar c^2}{(2\pi)^{3/2}} \frac{\sigma_k k^3 \theta_s^2}{e^{\hbar c k / k_\mathrm{B}T_s} -1},
\end{alignat}
with $k_\mathrm{B}$ the Boltzmann constant. The finite angular size of the source is taken into account by accompanying every factor of $\langle I_s \rangle$ in the numerator of \Eq{eq:C_epic} with a form factor $\mathcal{F}_s(y) \equiv 2 J_1(y)/y$ with $y \equiv (\theta_s/\sigma_{\theta_\mathrm{res}})$, i.e.~the 2D Fourier transform of a uniform disk at angular wavenumber $k\vect{d}$. 

The SNR on the intensity correlation in a single spectral channel is $C/\sigma_C$ for polarized light, and $C/2\sigma_C$ for unpolarized light. If one can disambiguate the fringe number of $\vect{\theta}_{ba}$ (Methods~\ref{app:obs}), the \emph{per-channel} light-centroiding precision becomes
\begin{alignat}{2}
  \sigma^{(1)}_{\delta \theta} &= \left(\frac{1}{2\sigma_C}\frac{\dd C}{\dd (\delta \theta_{ba})} \right)^{-1} \label{eq:sigma_channel} \\
  &= \sigma_{\theta_\mathrm{res}} \frac{\sigma_C}{\Big|\sin \Big(\frac{\delta \theta_{ba}}{\sigma_{\theta_\mathrm{res}}} \Big)\Big|} 
  \frac{4\sqrt{2} c \sigma_k \sigma_t}{\mathcal{F}_a\mathcal{F}_b}
  \frac{\left(\langle I_a \rangle +  \langle I_b \rangle\right)^2}{\langle I_a \rangle \langle I_b \rangle} \nonumber
\end{alignat}
by standard error propagation. The light-centroiding precision from the combination over all spectral channels labeled by $k$, with $n_\mathrm{arr}$ detectors per array site, is the inverse quadrature sum of \Eq{eq:sigma_channel}:
\begin{alignat}{2}
  \sigma_{\delta \theta} 
  &= \frac{1}{n_\mathrm{arr}} \left[ \sum_k  \left(\sigma_{\delta \theta}^{(1)}\right)^{-2} \right]^{-1/2} \label{eq:sigma_theta}\\
  &\simeq \frac{2^{13/2} \pi^{5/4} \hbar^3 c^3}{k_\mathrm{B}^3} \frac{1}{n_\mathrm{arr} A d} \sqrt{\frac{\sigma_t}{\eta^2 t_\text{obs}}\frac{1}{\mathcal{R}}} \frac{1}{T_s^3 \theta_s^2} \mathcal{I}^{-1/2}, \nonumber
\end{alignat}
where the sum runs over all spectral channels centered on wavenumbers $k = (2 \pi / \lambda_\mathrm{max}) e^{2m/\mathcal{R}}$ with $m = 0, 1, \dots, \lfloor (\mathcal{R}/2) \ln(\lambda_\mathrm{max}/\lambda_\mathrm{min}) \rfloor$ with minimum and maximum wavelengths, assumed to be $\lambda_\mathrm{min} = 300\,\mathrm{nm}$ and $\lambda_\mathrm{min} = 1{,}000\,\mathrm{nm}$ in \Tab{tab:phases} and \Fig{fig:exoplanets}. In the second line, we have evaluated and approximated this sum with an integral to give the parametric dependence on telescope properties (second fraction), detection specifications (square root), and source parameters (fourth fraction). The telescopes and detectors are assumed to be the same at both sites ($\eta = \eta_1 = \eta_2$, $A = \pi D^2 / 4 = A_1 = A_2$, etc.); likewise for the sources $s = a,b$, with identical $T_s$ and $\theta_s$. The final factor is the (inverse square root of the) dimensionless integral:
\begin{alignat}{2}
  \mathcal{I} \equiv \int_{x_\mathrm{min}}^{x_\mathrm{max}} \dd x \, \frac{x^5}{\big(e^{x}-1\big)^2} \left[\mathcal{F}_s\left(x \frac{k_\mathrm{B} T_s \theta_s d}{\hbar c} \right)\right]^4,
\end{alignat}
with $x_\mathrm{min} = 2\pi \hbar c / k_\mathrm{B} T_s \lambda_\mathrm{min}$ and similar for $x_\mathrm{max}$. The suppression of $\mathcal{F}_s$ and thus $\mathcal{I}$ at large $d$ is why there is an optimal baseline $d^\mathrm{opt}$ for differential light-centroiding. This optimal value depends on $\lambda_\mathrm{min}$, $\lambda_\mathrm{max}$, $T_s$, and $\theta_s$, but is roughly that for which $\sigma_{\theta_\mathrm{res}} \sim \theta_s$ in the most sensitive spectral channel. For a pair of Sun-like stars ($T_s = 6{,}000 \, \mathrm{K}$, $R_s = R_\odot$), this optimal baseline is
\begin{alignat}{2}
  d^\mathrm{opt} \approx 0.71\,\mathrm{km} \left(\frac{D_s}{100\,\mathrm{pc}}\right) \label{eq:d_opt}
\end{alignat}
for the assumed spectral range. For hotter stars even more suitable to EPIC, $d^\mathrm{opt}$ would be larger and $\sigma_{\theta_\mathrm{res}}$ better (at fixed intensity). Equation~\ref{eq:d_opt} sets the fiducial resolution and the other resulting angular scales in \Tab{tab:phases} and \Fig{fig:exoplanets}.

\section{Atmospheric Noise} \label{app:atm}
One of the main advantages of  traditional intensity interferometry, preserved by EPIC, is that the \emph{differential} light-centroiding precision $\sigma_{\delta \theta}$ is impervious to atmospheric aberrations for small source separations $\theta_{ba}$. Any fluctuation in the index of refraction $n[\vect{x}]$ will be \emph{common} to $a \to p$ and $b \to p$ for any $p$ separately, and will not contribute to the doubly-differential phase in the second line of \Eq{eq:C_epic} for the same reason that a common extension/delay will not alter \Eq{eq:path_epic}. We write the atmospheric phase fluctuation as
\begin{alignat}{2}
  \widetilde{\phi}_{sp} = k \int_0^{r_{sp}} \dd l \, n[\vect{x}_{sp}(l)] \label{eq:phi_atm}
\end{alignat}
with the mean refraction subtracted out: $\langle n[\vect{x}] \rangle = 0 ~ \forall ~ \vect{x}$ and thus $\langle \widetilde{\phi}_{sp} \rangle = 0$. Equation~\ref{eq:phi_atm} is a line-of-sight integral over the $p \to s$ path, namely $\vect{x}_{sp}(l) \equiv \vect{r}_p + l \, \hat{\vect{\theta}}_s$. 
At optical wavenumbers $k$, the atmospheric phase variance is enormous: $\langle \widetilde{\phi}_{sp}^2 \rangle \gg 10^4$.

Because of the small spatial coherence of the fluctuating index of refraction in the turbulent atmosphere, any intensity interferometric scheme where the light from $a \to p$ and $b \to p$ reach the same photodetector through \emph{separate} apertures will \emph{not} enjoy phase cancellation, thus erasing all fringe contrast. Between the inner scale $l_0 \sim 1 \, \mathrm{mm}$ and the outer scale $L_0 \sim 10 \, \mathrm{m}$, spatial fluctuations in the refractive index are statistically quantified by the structure function:
\begin{alignat}{2}
   \left \langle \left(n[\vect{x} + \vect{r}] - n[\vect{x}] \right)^2 \right \rangle = C_n^2[\vect{x}] \, r^{2/3}, 
\end{alignat}
valid for $l_0 \ll r \ll L_0$ and an overestimate elsewhere. The ``constant'' $C_n$ is only weakly dependent on position, and is mostly a (decreasing) function of altitude, with $C_n \sim \mathcal{O}(10^{-8} \, \mathrm{m}^{-1/3})$ at $1\,\mathrm{km}$ and $\mathcal{O}(10^{-9} \, \mathrm{m}^{-1/3})$ at $10\,\mathrm{km}$.

The \emph{differential} atmospheric phase variance 
\begin{alignat}{2}
  \sigma_{\widetilde{\phi},p}^2 \equiv \left\langle \left(\widetilde{\phi}_{bp} - \widetilde{\phi}_{ap} \right)^2 \right\rangle = \left(\frac{\theta_{ba}}{\theta_{0,p}}\right)^{5/3}
\end{alignat}
between the wavefronts of $a$ and $b$ from a single vantage point $p$ can be small, as long as the source separation $\theta_{ba}$ is much smaller than the isoplanatic angle:
\begin{alignat}{2}
  \theta_{0,p} \equiv \left[2.9 k^2 \int_0^{r_{sp}} \dd l \, C_n^2[\vect{x}_{sp}(l)] \, l^{5/3} \right]^{-3/5},
\end{alignat}
itself a smoothly varying function of the position of $p$ and the sources' angle from zenith. A further calculation~\cite{longpaper} reveals that the fringe in the second line of \Eq{eq:C_epic} is suppressed by the factor $\exp \lbrace - (\sigma_{\widetilde{\phi},1}^2 + \sigma_{\widetilde{\phi},2}^2)/2 \rbrace$. Intensity correlation fringe contrast remains essentially unaltered for sources within the same isoplanatic patch.

This analysis shows that while a source-dependent path extension could be introduced with a double aperture and a beam recombiner~\cite{2022OJAp....5E..16S,Chen:2022ccn}, such a setup would be susceptible to severe refractive phases from the turbulent atmosphere, negating one of the core advantages of intensity interferometry. To avoid atmospheric phase noise, it is imperative that the beams from both sources traverse the same air column down to millimeter accuracy.

The need for a ``nearly common beam'' necessitates the beam splitting of \Fig{fig:basics}(b) for wide-angle astrometry with EPIC, but makes possible exceptional differential astrometric measurements from ground-based observatories despite potentially poor atmospheric conditions.

\section*{Author Contributions}
KVT conceived the experimental setup and produced the figures. KVT, MB, MG, and NW worked out all theoretical and practical aspects of the technique and its scientific applications. All authors contributed to the manuscript.

\section*{Data availability}
All data and code are available from the authors upon reasonable request.

\section*{Competing Interests}
The authors declare no competing interests.


\end{document}